\documentclass{elsart3p-mod}

\usepackage{graphics}
\usepackage{graphicx}
\usepackage{amssymb}
\usepackage{upgreek}

\usepackage{lineno}

\newcommand{\comment}[1]{}

\begin{document}

\begin{frontmatter}



\title{Status and First Results of the Acoustic Detection Test System AMADEUS}


\author{Robert Lahmann\corauthref{rl} on behalf of the ANTARES Collaboration}
\corauth[rl]{robert.lahmann@physik.uni-erlangen.de}
%\ead{robert.lahmann@physik.uni-erlangen.de}
\address{Erlangen Centre for Astroparticle Physics (ECAP), Erwin-Rommel-Str.\ 1,
 91058 Erlangen, GERMANY}

\begin{abstract}

The AMADEUS system is integrated in the ANTARES neutrino telescope in
the Mediterranean Sea and aims for the investigation of
acoustic particle detection techniques in the deep sea. Installed at a depth of
more than 2000\,m, the acoustic sensors of AMADEUS are using piezo-ceramic
elements for the broad-band recording of acoustic signals with
frequencies ranging up to 125\,kHz.
AMADEUS consists of six clusters, each one comprising six acoustic sensors
that are arranged at distances of roughly 1\,m from each other. 
Three
acoustic clusters are installed along a vertical mechanical structure
(a so-called Line) of ANTARES with spacings of
about 15\,m and 110\,m, respectively. The remaining 3 clusters
are installed
with vertical spacings of 15\,m on a further Line of the ANTARES
detector. The horizontal distance between the two lines is 240\,m.
Each acoustic cluster allows for the suppression of random noise by
requiring local coincidences and the reconstruction of the arrival
direction of acoustic waves.  Source positions can then be
reconstructed using the precise time correlations between the clusters
provided by the ANTARES clock system.
AMADEUS thus allows for extensive acoustic
background studies including signal correlations on several length
scales as well as source localisation. The system is therefore
excellently suited for feasibility studies for a potential future large scale
acoustic neutrino telescope in sea water.
Since the start of data taking on December 5th, 2007 a wealth of data
has been recorded. The AMADEUS system will be described and some first
results will be presented.

\end{abstract}

\begin{keyword}
AMADEUS \sep ANTARES \sep neutrino telescope \sep acoustic neutrino detection 
\sep thermo-acoustic model 

\PACS 95.55.Vj \sep 95.85.Ry \sep 13.15.+g \sep 43.30.+m
\end{keyword}
\end{frontmatter}

% main text

\section{Introduction}
The production of pressure waves by cascades of fast particles passing
through liquids was predicted as early as 1957~\cite{Askariyan1},
leading to the development of the so-called \emph{thermo-acoustic
  model} in the 1970s~\cite{Askariyan2,Learned}.
According to the model, the energy deposition of charged particles in
liquids leads to a local heating of the medium which can be regarded
as instantaneous with respect to the typical time scale of the
acoustic signals.  Because of the temperature change the medium
expands or contracts according to its volume expansion coefficient
$\alpha$.  The accelerated motion of the heated medium produces a
pressure pulse which propagates through the medium.

\begin{figure}[ht]
\centering
\includegraphics[width=7.6cm]{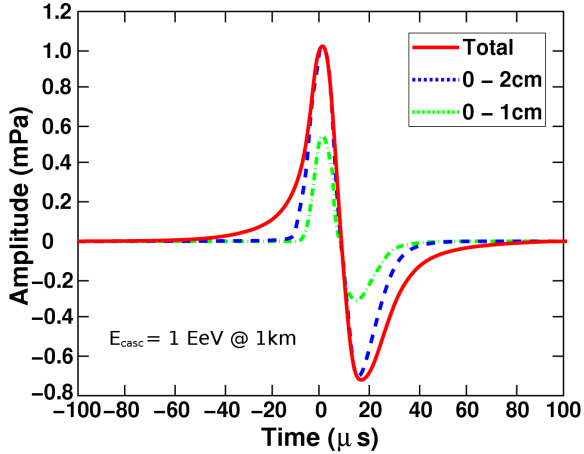}
\caption{ Simulated bipolar pressure pulse from a $10^{18}$\,eV
  hadronic cascade at a radial distance of 1\,km.  Roughly half of the
  pressure pulse is produced within a radius of 1\,cm (dash-pointed
  line) of the cascade, whereas the energy distribution within a
  radius of 2\,cm (dashed line) is nearly completely responsible for
  the final signal shape (solid line).  Figure adapted
  from~\protect\cite{bib:Acorne2007}.  }
\label{fig:bipolar_pulse}
\end{figure}

Fig.~\ref{fig:bipolar_pulse}
shows a bipolar pressure pulse from a recent Monte Carlo simulation of
neutrino interactions in water for a hadronic cascade with an energy of
1\,EeV~\cite{bib:Acorne2007}.  The amplitude was calculated at a
distance of 1\,km from the cascade in a plane perpendicular to the
shower axis at the shower maximum.

The cascades extend over a length of a few metres with a radius of
only a few cm. In radial direction, i.e. perpendicular to the shower
axis, the coherent superposition of the emitted sound waves leads to
the propagation of the sound within a flat disk (often referred to as
{\em pancake}) with a small opening angle.  %Assuming 
With a linear energy 
dependence, the pressure pulse amplitude scales as
\begin{equation}
p(\mathrm{1\,km}) \approx  1\,\mathrm{mPa} \cdot \frac{E_\mathrm{casc}}{1\,\mathrm{EeV}} 
\end{equation}
with the spectral energy density peaking around 10\,kHz.

Two major advantages over an optical neutrino telescope make acoustic
detection worth studying.  First, the attenuation length is on the
order of 5\,km (1\,km) for 10\,kHz (20\,kHz) signals.  This is
approximately one order of magnitude higher than for Cherenkov light
(attenuation length 60\,m or below) in the relevant frequency band.

The second advantage is the
much simpler sensor design and readout electronics required for acoustic
measurements:  
No high voltage is required in the case of acoustic measurements
and time scales are in the 
$\upmu$s range for acoustics as compared to the ns range for optics. This
allows for online implementation of advanced signal processing
techniques. Efficient data filters are essential, as the signal
amplitude is relatively small compared to the acoustic background in
the sea, which complicates the unambiguous determination of the
signal.

Understanding the background conditions at the site of a potential
large-scale acoustic neutrino detector is therefore crucial in
order to assess the feasibility of such a device.  It is important to
realize that there are two kinds of background which need to be
understood: First, there is ambient noise which can be described by
its characteristic power spectral density.  This noise is determined
by environmental processes and will be discussed in
Sec.~\ref{sec:first-results}.  This background defines the minimum
pulse heights that can be measured if a given signal-to-noise ratio
can be achieved with a search algorithm.  To measure this background,
in principle one hydrophone is sufficient and the synchronisation among
multiple hydrophones is not crucial.

Second, there are neutrino-like events, i.e. signals which have the 
characteristic bipolar pulse shape shown in Fig.~\ref{fig:bipolar_pulse}, 
but have a different origin. 
It is important to measure their spatial, temporal and pulse-height 
distribution in order to assess the probability for random coincidences that
mimic the characteristic pancake structure of a neutrino sound wave. 
For this kind of measurement, a hydrophone array is required and the 
synchronisation among the hydrophones is crucial.

A huge number of signals recorded by AMADEUS are emitted
by marine mammal, mainly dolphins,
which is of great interest to marine biologists. 
The prospective research activities in this area will however not
be addressed in this paper. 

In the following section, an overview of the AMADEUS System will be
given (Sec.~\ref{sec:acoustics}), which will be followed by a
description of the components of the system in
Sec.~\ref{sec:components-AMADEUS}.  The system response as measured in
the laboratory prior to the deployment will be discussed in
Sec.~\ref{sec:sys-response} and some first results will be discussed
in Sec.~\ref{sec:first-results}.

\section{Overview of the AMADEUS System}
\label{sec:acoustics}

\subsection{Goals}
\label{subsec:goals}
It is the declared main goal of the AMADEUS project to perform a
feasibility study for a potential future large-scale acoustic neutrino
detector. To this end, the following aims are pursued:
\begin{itemize}
\item
Long term background investigations 
(rate of neutrino-like signals, localisation of sources, levels of 
ambient noise);
\item
Investigation of signal and background correlations on different
length scales;
\item
Development and tests of filter and reconstruction algorithms;
\item
Tests of different sensors and sensing methods;
\item
Studies of hybrid detection methods.
\end{itemize}
These goals were driving the design of the AMADEUS system, which
will now be discussed.

\subsection{AMADEUS as Part of ANTARES}
\label{sec:amadeus_part_of_antares}

\begin{figure}[bht]
\centering
\includegraphics[width=7.6cm]{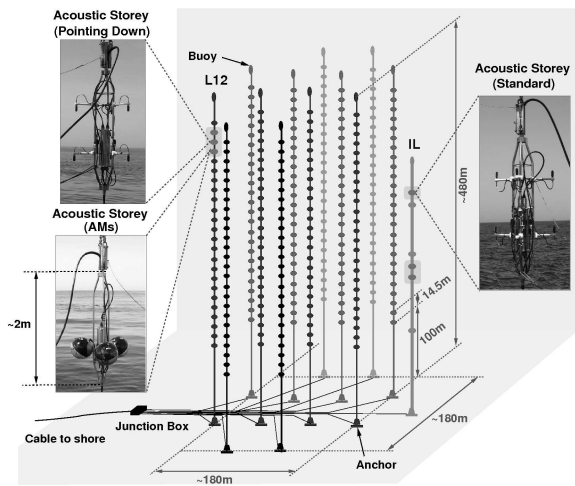}
\caption{A sketch of the ANTARES detector. 
The six Acoustic Storeys are highlighted and their three different setups
are shown. 
}
\label{fig:ANTARES_schematic_all_storeys}
\end{figure}

AMADEUS is a part of the ANTARES neutrino telescope~\cite{bib:ANTARES}
in the Mediterranean Sea, which was completed on May 30th, 2008.  A
sketch of the detector is shown in
Fig.~\ref{fig:ANTARES_schematic_all_storeys}.  ANTARES is located
in the deep sea, about 40\,km south of Toulon, 
at a water depth of about 2500\,m. It comprises 12 vertical
structures, the {\em detection lines} (or lines for short) plus a 13th
line, called {\em Instrumentation Line} or {\em IL}, which is equipped
with instruments for monitoring the environment of the detector.  Each
detection line holds 25 {\em storeys} that are arranged at equal
distances of 14.5\,m along the line, interlinked by
electro-mechanical-optical cables.  A standard storey consists of a
titanium support structure, holding three {\em Optical Modules (OMs)},
i.e. photomultiplier tubes (PMTs) inside glass spheres with a nominal
diameter of 432\,mm, and one {\em Local Control Module (LCM).} The LCM
holds the offshore electronics and the power supply within a
cylindrical titanium container (cf. Sec.~\ref{sec:offshore}).
Each line is fixed on the sea floor by an anchor ({\em Bottom String
Socket, BSS}) and held vertically by a buoy.

Acoustic sensing was integrated in form of {\em Acoustic Storeys}
which are modified versions of standard ANTARES storeys: Acoustic
sensors replace the OMs and custom-designed electronics is used to
digitise and preprocess the analogue signals. Details
will be discussed below.

The three Acoustic Storeys on the IL started data taking when the
connection to shore was established on Dec. 5th, 2007. The
Acoustic Storeys on Line 12 (L12) were connected to shore during the
completion of ANTARES in May 2008.  AMADEUS is now fully functional
with 34 of its 36 sensors working.

\subsection{Description of the System}

It has been a fundamental design principle of the AMADEUS system to
make use of standard ANTARES hard- and software as much as possible.
The intention was to minimise the effort for design and engineering
and to reduce the failure risks for both ANTARES and AMADEUS by
reducing the need for additional quality assurance and control measures
to a minimum.
In order to integrate the system successfully into the ANTARES
detector, design efforts in three basic areas were necessary: First,
the development of acoustic sensors that replace the OMs of standard
ANTARES storeys; second, the development of an offshore acoustic ADC
and preprocessing electronics board; third, the development of on- and
offline software.
These subjects will be discussed in more detail in
Sections \ref{sec:acousensors},
\ref{sec:acouadc-board} and \ref{sec:onshore},
respectively.

\begin{figure}[ht]
\centering
\includegraphics[height=7.2cm]{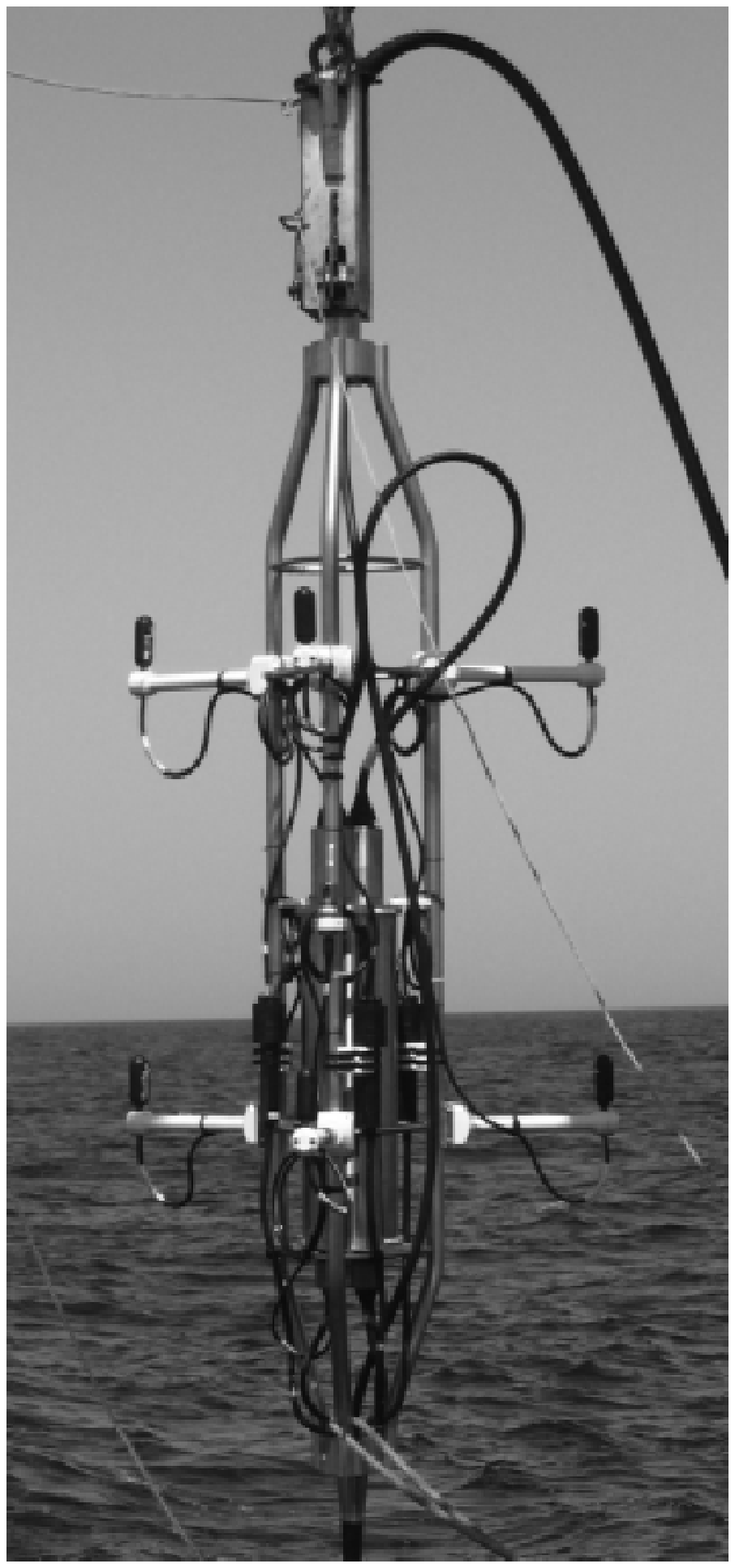}
\hspace{0.75mm}
\includegraphics[height=7.2cm]{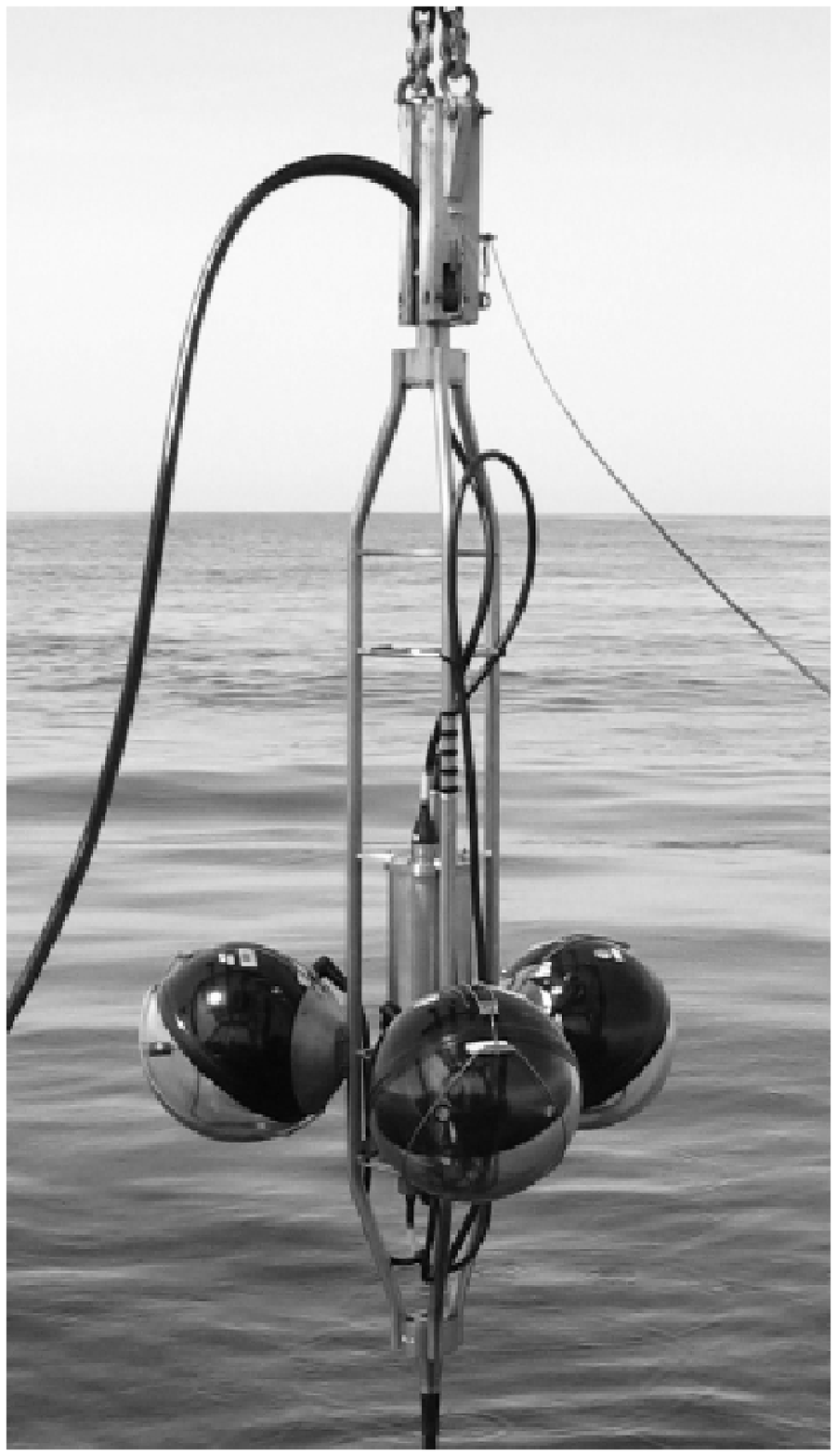}
\caption{An Acoustic Storey of the IL (left), equipped with
  hydrophones, and the lowermost Acoustic Storey of L12 (right)
  equipped with Acoustic Modules, shown during their deployment.}
\label{fig:antares_storey_acou}
\end{figure}

Each Acoustic Storey is equipped with six acoustic sensors with
interspacings on the order of 1\,m.  The Acoustic Storeys on the IL
are located at 180\,m, 195\,m, and 305\,m above the sea floor,
respectively.  Line 12 is anchored at a horizontal distance of about
240\,m from the IL, with the Acoustic Storeys positioned at heights of
380\,m, 395\,m, and 410\,m.  With this setup, the maximum distance
between two Acoustic Storeys is 340\,m.  Two of the six Acoustic
Storeys are shown in Fig.~\ref{fig:antares_storey_acou}.

The final system has the full capabilities of a detector, such as time
synchronisation and a 
continuously operating data acquisition, and is scalable to a larger
number of Acoustic Storeys.
It combines local clusters of acoustic sensors (the Acoustic Storeys)
with large cluster spacings, allowing for fast direction
reconstruction with individual storeys that then can be combined to
reconstruct the position of a source.

\section{Components of the AMADEUS System}
\label{sec:components-AMADEUS}
\subsection{The Acoustic Sensors}
\label{sec:acousensors}

Two types of acoustic sensors are used in AMADEUS: hydrophones and
so-called {\em Acoustic Modules} ({AMs},
cf. Fig.~\ref{fig:antares_storey_acou}). In both cases, the sensors
are based on piezo-electrical ceramic elements that convert pressure waves
into voltage signals, which are then amplified for
readout~\cite{bib:hoessl2006}. The ceramics and amplifiers are coated
in polymer plastics in the case of the hydrophones. Those sensors
have a diameter of about 4\,cm and a length (from the cable junction
to the opposite end) of about 10\,cm. 

For the AMs, the ceramics are glued to the
inside of the same spheres used for the Optical Modules of
ANTARES. This design was inspired by the idea
to investigate an option for acoustic sensing that can be combined
with a PMT in the same housing.

In order to obtain a complete 2$\pi$-coverage of the azimuthal angle
$\phi$, the 6 sensors are distributed over the 3 AMs of the storey
within a plane perpendicular to the longitudinal axis of a storey.
The two piezo elements within one sphere are arranged at an opening
angle of 60$^\circ$.\footnote{ For reasons such as limited data rate
  and the pre-existing design of the ANTARES offshore electronics,
  implementing more than 2 sensors per AM would have increased the
  technical effort disproportionately.}

The three Acoustic Storeys on the IL house hydrophones only, whereas
the lowermost storey of Line 12 holds AMs
(cf. Fig.~\ref{fig:ANTARES_schematic_all_storeys}).  The standard
mounting of the hydrophones is with their cable junction pointing
downwards. In one of the storeys on Line 12, the hydrophones were
mounted with the cable junction, where the sensitivity is largely
reduced, pointing upwards.  This allows for
investigations of the directionality of background from ambient noise,
which is expected to come mainly from the sea surface.

Three of the five storeys holding hydrophones are equipped with
commercial hydrophones\footnote{Custom produced by High
  Tech Inc (HTI) in Gulfport, MS (USA). 
}  and the other two with hydrophones developed and
produced at the Erlangen Centre for Astroparticle Physics (ECAP).

All acoustic sensors are tuned to have a low noise level and to be
sensitive over the whole frequency range of interest from 1 to 50\,kHz.
Their typical sensitivity is around $-145$\,dB\,re.\,1V/$\upmu$Pa
(including preamplifier) \cite{bib:naumann_phd}.
\subsection{Offshore Electronics and Acoustic Data Acquisition}
\label{sec:offshore}

In the ANTARES DAQ scheme~\cite{bib:antares_daq}, the digitisation 
of sensor signals is
conducted within the offshore electronics container on each storey by
several custom-designed electronics boards, which send all digitised
data to shore. Here data reduction is performed and the data is
searched for events of interest.  With its capability of timing
synchronisation on the nanosecond-scale\footnote{ The clock system is in
  fact capable of providing sub-nanosecond precision for the
  synchronisation of the optical data recorded by the PMTs. This
  precision, however, is not required for the acoustic data.}  and
transmission of several MByte per second and per storey, it is
perfectly suited for the acquisition of acoustic data. In addition, in
each LCM a {\em Compass board} measures the tilt and the orientation
of the storey.

For the digitisation and preprocessing of the acoustic signals and 
for feeding them into the ANTARES data stream, 
the {\it AcouADC board} was designed.
Each board processes the differential signals from two acoustic 
sensors, which results in a total of three such boards
per storey.

\begin{figure}[ht]
\centering
\includegraphics[width=7.7cm]{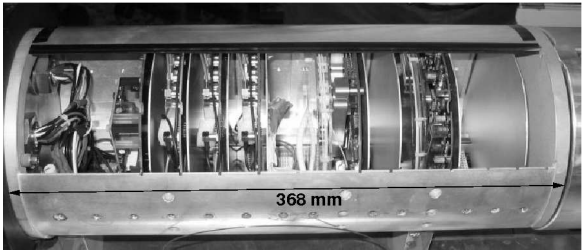}
\hspace{2pc}%
  \caption{ An LCM crate equipped with AcouADC boards before insertion
    into its titanium housing.  From the left to the right, the
    following boards are installed: The Compass board; 3 AcouADC
    boards; a {\em Data Acquisition (DAQ) board} that sends the data
    to shore; and a {\em Clock board} that provides the timing signals
    to correlate measurements performed in different storeys.
\label{fig_Acou_LCM}}
\end{figure}

Figure~\ref{fig_Acou_LCM} shows the fully equipped LCM of an Acoustic
Storey.  
All components except for the  AcouADC boards are 
standard components that are present in every LCM crate of ANTARES. 

\subsection{The AcouADC Board}
\label{sec:acouadc-board}
\begin{figure}[ht]
\centering
\includegraphics[width=5.5cm]{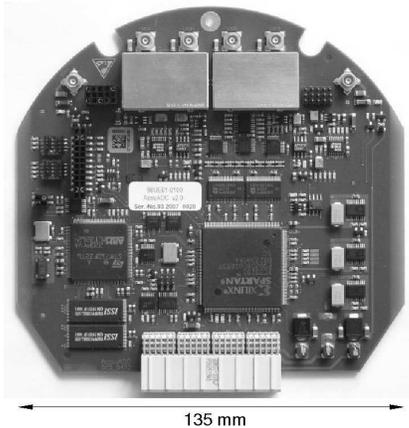}\hspace{2pc}%
  \caption{
  An AcouADC board. 
\label{fig_AcouADC_board}}
\end{figure}

The AcouADC board is shown in Fig.~\ref{fig_AcouADC_board}.  It
consists of an analogue and a digital part.  The analogue part
amplifies the voltage signals coming from the acoustic sensors by one
of 12 adjustable factors between 1 and 562 and applies a bandpass
filter to the resulting signal.  To protect the analogue part from
potential electromagnetic interference, it is shielded by metal
covers.  The system has low noise and is designed to be -- together
with the sensors -- sensitive to the acoustic background of the deep
sea over a wide frequency-range.  
The dynamic range achieved for the standard gain factor of 10 is from
about 5\,mPa to 5\,Pa in peak-to-peak amplitude of an acoustics signal
over the frequency range from 1 to 100\,kHz.

The analogue filter 
suppresses frequencies below 4\,kHz and above 
130\,kHz. The high-pass part cuts into the trailing edge of the low
frequency noise of the deep-sea acoustic background~\cite{urick} and
thus protects the system from saturation. The low-pass part
efficiently suppresses frequencies above the Nyquist frequency of
250\,kHz for the maximum sampling rate of 500\,kSamples per second (kSPS). 
Within the passband, the filter response is essentially flat with a linear 
phase response.

The digital part of the AcouADC board digitises and processes the
acoustic data. It is designed to be highly flexible by employing a
micro controller ($\upmu$C) and a field programmable gate array (FPGA)
as data processor. The $\upmu$C can be controlled with the onshore
control software and is used to adjust settings of the analogue part
and the data processing.  Furthermore, the $\upmu$C can be used to
update the firmware of the FPGA in situ.

The digitisation is done by one 16-bit ADCs for each of the two input
channels.  The digitised data from the two channels is read out in
parallel by the FPGA and further processed for transmission to the DAQ
board.  The DAQ board then handles the transmission of the data to the
onshore data processing servers.

In standard mode, the digitised data is down-sampled to 250\,kSPS by
the FPGA, corresponding to a downsampling by a factor of 2.  Hence the
frequency spectrum of interest from 1 to 100\,kHz is fully contained
in the data.

Onshore a dedicated computer cluster is used to process and store
the acoustic data arriving from the storeys and to control the offshore
DAQ. This is discussed below.

\subsection{Onshore Data Processing}
\label{sec:onshore}
AMADEUS follows the same ``all data to shore'' strategy as ANTARES; 
the offshore data arrives via the TCP/IP protocol at a Gigabit switch 
in the ANTARES control room, where the acoustic data is separated from 
the standard ANTARES data and routed to the acoustic computer cluster.

The cluster currently consists of four servers of which two are used for 
online data filtering. The filtering has the task of 
reducing the raw data rate of about 1.5~TB/day to about ~15 GB/day for 
storage. The filter schemes that are currently implemented 
are described in~\cite{Neff_these_procs}.
Furthermore, all components are scalable which makes 
the system extremely flexible. Additional servers
can be added or the existing ones can be replaced by the latest models
if more sophisticated filter algorithms are to be implemented.
In principle it is also possible to move parts of the filter into the FPGA
of the AcouADC board, thereby implementing an offshore trigger which reduces
the size of the data stream sent to shore.

Just like ANTARES, AMADEUS can be controlled via the Internet from 
essentially any place in the world and is currently operated from Erlangen. 
Data are centrally stored and are available remotely as well.

\section{The System Response}
\label{sec:sys-response}
Every individual component of the complete data taking chain was
calibrated in the laboratory prior to deployment.

For each hydrophone and each channel of the AMs, the sensitivity was measured
up to 100\,kHz as a function of the azimuthal and polar angle.

\begin{figure}[tbh]
\centering
\includegraphics[height=7.0cm]{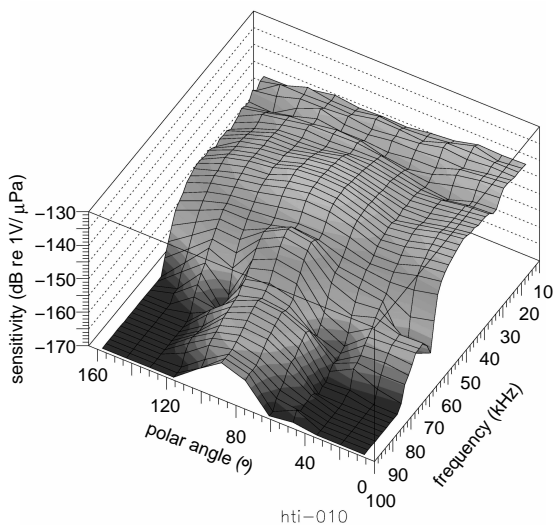}
\caption{
Sensitivity of a typical AMADEUS hydrophone 
as a function of frequency and polar angle.
$0^\circ$ corresponds to the direction opposite to the cable junction. 
}
\label{fig:hydrophone_sensitivity}
\end{figure}

The sensitivity measurements were done with a calibrated sender
stimulated by a Gaussian voltage pulse. The signal recorded by the
acoustic sensor and the stimulating voltage pulse were transformed into
the frequency domain, divided and corrected for the sender
characteristics.
As a result, the corrected sensitivity in dB re 1V/$\upmu$Pa is
obtained as a function of frequency for a given orientation of the
sensor with respect to the sender.

For a given frequency, the sensitivity for all acoustic sensors is
essentially flat as a function of the azimuthal angle on a 3\,dB
level.
The sensitivity as a function of frequency and polar angle
for a typical hydrophone is shown in
Fig.~\ref{fig:hydrophone_sensitivity} ~\cite{bib:naumann_phd}.  The
highest sensitivity is recorded in the direction
perpendicular to the longitudinal axis of the hydrophone,
i.e. in the horizontal plane within an Acoustic Storey.

Furthermore, 
using various input stimuli at the differential analogue input
connectors of the AcouADC board, the transfer function of the complete chain
consisting of the electronic components forming the analogue filter and 
amplification stages was measured and parameterised.
The digital filter that is applied in case of downsampling adds
an additional, frequency-independent time delay. 

\begin{figure}[ht]
\centering
\includegraphics[width=7.8cm]{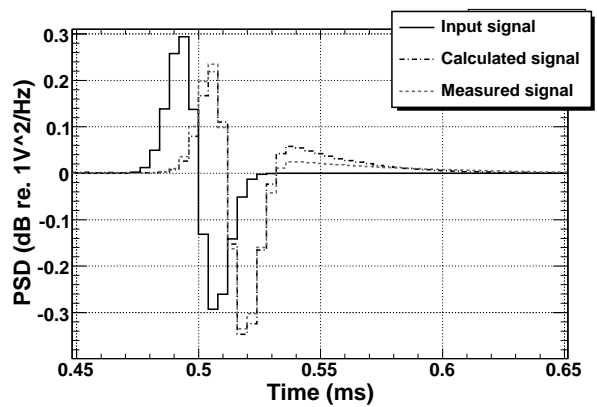}
\caption{A neutrino-like bipolar input signal to one of the analogue
  inputs of an AcouADC board, the signal recorded at the
  input of the ADC and the expected signal according to the
  parameterised transfer function.  The measurement was done for a
  nominal gain factor of 1. }
\label{fig:signal_response_acouADC}
\end{figure}

In Fig.~\ref{fig:signal_response_acouADC}, a bipolar
input signal is compared to the signal measured at the input of the
ADC and to the result of the parameterisation of the system
behaviour. The agreement is very good.

The gains of the analogue amplification were calibrated and found to
agree to better than 10\% with their nominal values. The system
stability was checked both in the laboratory and in situ and was found
to be excellent. All further relevant characteristics of the ADCs were
investigated and are in full compliance with the requirements.

\section{First Results}
\label{sec:first-results}
\subsection{Ambient Noise Measurements}
The ambient noise level in the frequency range from about 200\,Hz to 50\,kHz 
in the deep sea is assumed to be mainly determined
by the agitation of the sea surface, i.e. by waves, spray and 
precipitation~\cite{urick2}.
To verify these assumptions, the correlation between the weather and
the RMS of the ambient noise recorded by AMADEUS was investigated.
Weather data at the Hy\`eres airport at the French coast, about 30\,km north of 
the ANTARES site, is continuously logged by the AMADEUS onshore software.
New data is made available every hour.

\begin{figure}[ht]
\centering
\includegraphics[width=7.8cm]{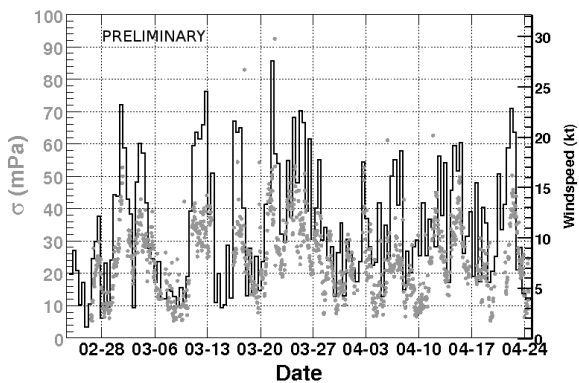}
\caption{RMS noise recorded by a 
representative hydrophone (dots) and wind speed (solid line) as a 
function of time, showing the correlation between the two quantities.
Wind speed is given in knots.}
\label{fig:correlation_wind_rms_phd}
\end{figure}

An upgrade of the study performed in~\cite{bib:Graf_PhD_2008}
is shown in Fig.~\ref{fig:correlation_wind_rms_phd}.
Each point in the figure represents the RMS noise of a 10\,s sample of data, 
calculated by integrating the power spectral density (PSD) 
in the frequency range from 1 to 50\,kHz.
The correlation between the wind speed and the hydrophone noise is
found to be about 80\%.
The results are consistent with those reported for other deep-sea
sites~\cite{SAUND2008, ONDE2008}.

\begin{figure}[ht]
\centering
\includegraphics[width=7.8cm]{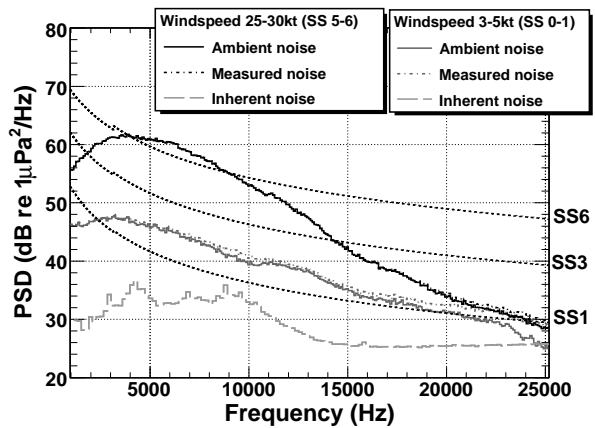}
\caption{Power Spectral Density (PSD) as measured for two time slices
  of 10\,s length for two different conditions of the sea (sea states
  (SS) 0 to 1 and 5 to 6, respectively) with a representative
  hydrophone. Shown are also the Knudsen curves~\protect\cite{urick} for 3
  different sea states, which correspond to the expected PSDs.  The
  ambient noise is the difference of the noise recorded in the deep
  sea and the inherent noise of the system, measured prior to
  deployment.  Wind speed is given in knots.}
\label{fig:noise_comp_arena}
\end{figure}

Furthermore, the ambient noise was investigated for different sea
states corresponding to different wind speeds as shown in
Fig.~\ref{fig:noise_comp_arena}.  The ambient noise, characterised by
the PSD, can be clearly seen to increase with the sea state.  Shown is
also the expected behaviour, as described by the Knudsen
curves~\cite{urick}, which however were measured in shallow water. For
the deep sea data from AMADEUS, one observes a stronger roll-off of
the PSD with increasing frequency than predicted.  This is
qualitatively in agreement with the measurements presented
in~\cite{SAUND2008}.

The decreasing of the PSD for frequencies smaller than about 4\,kHz
cannot be unambiguously interpreted at the time of the writing of this
document. It is possibly caused by effects stemming from the
preliminary calibration of the hydrophone or the weighting of the
directional sensitivity used to calculate the overall sensor
sensitivity.

\subsection{Positioning of Acoustic Storeys}
A further important measurement with the AMADEUS system is the
accurate determination of the relative positions of the Acoustic
Storeys within the ANTARES detector ({\em positioning}).  The ANTARES
positioning system~\cite{bib:ardid_VLVnT_procs} uses transceivers
(so-called pingers) at the BSS
(cf. Sec.~\ref{sec:amadeus_part_of_antares}) of each line in
combination with 5 acoustic receivers (positioning hydrophones)
arranged along each detection line.  The pingers emit tone bursts at 9
well-defined frequencies between 44\,522\,Hz and 60\,235\,Hz which can
be used by AMADEUS to determine the positions of the Acoustic Storeys
by means of triangulation.

With 6 hydrophones, a complete reconstruction (position and three
angles) of each Acoustic Storey can be done using the pinger signals.
This is not only an interesting task on its own, but is in fact
important for an accurate reconstruction of the positions of unknown
sources as will be described below.  The data recorded by the Compass
boards (cf.~Sec.~\ref{sec:offshore}) allow for cross checks of the
results.

\begin{figure}[ht]
\centering
\includegraphics[width=7.7cm]{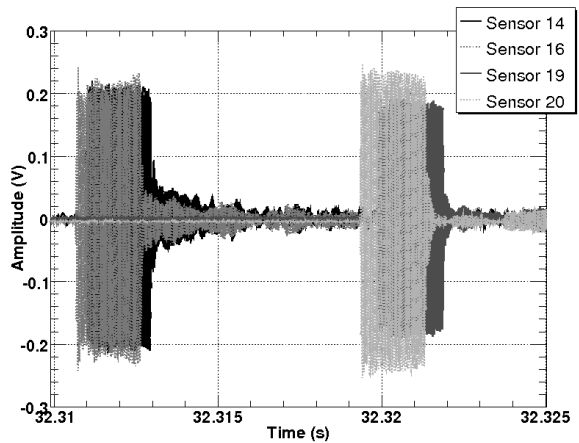}
\caption{Pinger signals received by four sensors in two different Acoustic 
Storeys on Line 12. 
The sensors in the lowermost Acoustic Storey (holding AMs, sensors
14 and 16) receive the signal earlier than those in the storey above
(holding hydrophones, sensors 19 and 20).  
\label{pinger_hit_am_full}
}
\end{figure}

Figure~\ref{pinger_hit_am_full} shows the signal emitted by one
pinger as recorded by two sensors in the AMs of the lowermost storey
and by two hydrophones of the central storey of Line 12.  One can
clearly observe the different arrival times of the signal
(corresponding to different travel times of the sound from the pinger)
between the storeys.
For the two sensors of each storey shown as an example, the smaller
differences in arrival times are also clearly visible.

Work on two methods for the positioning is currently in progress:
First, the differences between the absolute time of signal emission
and reception from several pingers are used to reconstruct the
position of each acoustic sensor individually.
Second, only the differences in arrival times of a pinger signal in
the 6 acoustics sensors of a storey are used to reconstruct the
direction of a pinger signal.  Then the reconstructed directions are
matched with the pinger pattern on the sea floor.
The second method employs the same algorithms used for position
reconstruction of unknown sources. %this will be described now.

\subsection{Position Reconstruction of Sources}

Position reconstruction of acoustic point sources will be done by first
reconstructing their direction from individual storeys and then
combining the reconstructed directions from three or more storeys.

To find the direction of point sources, a beam forming algorithm is
used.  It is designed to reconstruct plane waves, which for the
geometry of a storey is a reasonable assumption for sources with
distances $\gtrsim 100\,$m.  Details can be found
in~\cite{richardt_these_proc}.

\section{Conclusions and Outlook}
The AMADEUS system, which is dedicated to the investigation of
acoustic neutrino detection techniques, has been successfully 
installed and operated within the ANTARES detector.
Except for its small size, the system has all features required for
an acoustic neutrino telescope and hence is excellently suited for 
a feasibility study of a potential future large-scale acoustic 
neutrino telescope.

AMADEUS can be used as a multi purpose device for studies of neutrino
detection techniques, position reconstruction, and marine research.

First results were presented which demonstrate the potential of
AMADEUS, covering the measurement of ambient noise, positioning of the
acoustic storeys and position reconstruction of sources.

\section{Acknowledgements}
This study was supported by the German government through BMBF grant
05CN5WE1/7.  The author wishes to thank the organisers of the ARENA
2008 workshop for their efforts.

\end{document}